\PassOptionsToPackage{unicode}{hyperref}
\PassOptionsToPackage{hyphens}{url}
\documentclass[
]{scrartcl}
\usepackage{amsmath,amssymb}
\usepackage{iftex}
\ifPDFTeX
  \usepackage[T1]{fontenc}
  \usepackage[utf8]{inputenc}
  \usepackage{textcomp} 
\else 
  \usepackage{unicode-math} 
  \defaultfontfeatures{Scale=MatchLowercase}
  \defaultfontfeatures[\rmfamily]{Ligatures=TeX,Scale=1}
\fi
\usepackage{lmodern}
\ifPDFTeX\else
\fi
\IfFileExists{upquote.sty}{\usepackage{upquote}}{}
\IfFileExists{microtype.sty}{
  \usepackage[]{microtype}
  \UseMicrotypeSet[protrusion]{basicmath} 
}{}
\makeatletter
\@ifundefined{KOMAClassName}{
  \IfFileExists{parskip.sty}{%
    \usepackage{parskip}
  }{
    \setlength{\parindent}{0pt}
    \setlength{\parskip}{6pt plus 2pt minus 1pt}}
}{
  \KOMAoptions{parskip=half}}
\makeatother
\usepackage{xcolor}
\usepackage{longtable,booktabs,array}
\usepackage{calc} 
\usepackage{etoolbox}
\makeatletter
\patchcmd\longtable{\par}{\if@noskipsec\mbox{}\fi\par}{}{}
\makeatother
\IfFileExists{footnotehyper.sty}{\usepackage{footnotehyper}}{\usepackage{footnote}}
\makesavenoteenv{longtable}
\usepackage{graphicx}
\makeatletter
\def\maxwidth{\ifdim\Gin@nat@width>\linewidth\linewidth\else\Gin@nat@width\fi}
\def\maxheight{\ifdim\Gin@nat@height>\textheight\textheight\else\Gin@nat@height\fi}
\makeatother
\setkeys{Gin}{width=\maxwidth,height=\maxheight,keepaspectratio}
\makeatletter
\def\fps@figure{htbp}
\makeatother
\ifLuaTeX
  \usepackage{luacolor}
  \usepackage[soul]{lua-ul}
\else
  \usepackage{soul}
\fi
\usepackage[dvipsnames]{xcolor}
\usepackage{fancyhdr}
\PassOptionsToPackage{hyphens}{url}\usepackage{hyperref}

\AtBeginDocument{%
\author{%
Samantha Shorey\ (\texttt{sshorey@utexas.edu})\textsuperscript{1}\thanks{Corresponding author.}\and
Benjamin Mako Hill\ (\texttt{makohill@uw.edu})\textsuperscript{2}\and
Samuel C.\ Woolley\ (\texttt{swoolley@utexas.edu})\textsuperscript{1}\\[0.5em]
\textsuperscript{1}University of Texas at Austin\\
\textsuperscript{2}University of Washington%
}%
}
\ifLuaTeX
  \usepackage{selnolig}  
\fi
\usepackage{bookmark}
\IfFileExists{xurl.sty}{\usepackage{xurl}}{} 
\hypersetup{
  pdftitle={From Hanging Out to Figuring It Out},
  pdfauthor={Samantha Shorey; Benjamin Mako Hill; Samuel C. Woolley},
  pdfkeywords={participatory media, computational
thinking, commenting, learning, scratch, social
media, youth, children, education, programming},
  hidelinks,
  pdfcreator={LaTeX via pandoc}}

\title{From Hanging Out to Figuring It Out}
\usepackage{etoolbox}
\makeatletter
\providecommand{\subtitle}[1]{
  \apptocmd{\@title}{\par {\large #1 \par}}{}{}
}
\makeatother
\subtitle{Socializing Online as a Pathway to Computational Thinking}
\author{Samantha Shorey \and Benjamin Mako Hill \and Samuel C. Woolley}
\date{}

\begin{document}
\maketitle
\begin{abstract}
Although socializing is a powerful driver of youth engagement online,
platforms struggle to leverage engagement to promote learning. We seek
to understand this dynamic using a multi-stage analysis of over 14,000
comments on Scratch, an online platform designed to support learning
about programming. First, we inductively develop the concept of
``participatory debugging''---a practice through which users learn
through collaborative technical troubleshooting. Second, we use a
content analysis to establish how common the practice is on Scratch.
Third, we conduct a qualitative analysis of user activity over time and
identify three factors that serve as social antecedents of participatory
debugging: (1) sustained community, (2) identifiable problems, and (3)
what we call ``topic porousness'' to describe conversations that are
able to span multiple topics. We integrate these findings in a
theoretical framework that highlights a productive tension between the
desire to promote learning and the interest-driven sub-communities that
drive user engagement in many new media environments.
\end{abstract}

\section{Introduction}\label{introduction}

As programmers, and as people, we rarely get it right the first time.
Failure is part of building things. When faced with challenges, we often
seek the mentorship and guidance of knowledgeable others. In our
everyday lives, social connections act as a pathway for problem
solving---for teaching us how to build and how to build better.
Supporting the development of problem-solving skills has become a
priority for educators hoping to equip students with tools for the
modern world. New educational programs are focused on the development of
problem-solving strategies, such as working together and asking
questions---skills that are necessarily linked to interaction with other
people (Partnership for 21st Century Skills, 2019). Although the
collaborative nature of problem solving is implicit in much of the
research of digital media and learning, only recently has a body of work
on ``connected learning'' tied this social behavior to young people's
learning of higher-order thinking skills (Ito et al, 2013).

This study contributes to the emerging agenda of connected learning by
taking a three-stage approach to studying the relationship between
learning and commenting among users on Scratch---an online community
with millions of users that is arguably the most popular setting for
young people to learn block-based programming. We use grounded theory
analyses of 640 projects and 53 user project histories, as well as a
content analysis of an additional 600 projects, to examine and
articulate the learning-oriented benefits associated with social
interaction. We propose the construct of \emph{participatory debugging},
a process through which users on Scratch (``Scratchers'') leverage the
Comments space within projects to troubleshoot issues associated with
project design and to develop higher-order computational thinking
skills. We also identify three contextual factors that we argue are
uniquely important to the development and continued practice of
participatory debugging: (1) a sustained community, (2) identifiable
problems, and (3) low topic density, or what we call ``topic
porousness.'' We suggest that the third proposition is especially
significant because it articulates a central tension in connected
learning: the trade-off between the desire to promote learning of
particular concepts (like those associated with computational thinking)
and engagement in interest-driven sub-communities that Ito et
al.~(2009b) and others have recognized as a pathway for meaningful
learning. We conclude by synthesizing our findings and exploring their
implications for the design of connected learning in various contexts.

\section{Background}\label{background}

\subsection{Socializing and Learning}\label{socializing-and-learning}

Influential approaches to education suggest that learning practices are
strongly situated in social environments (Lave and Wenger, 1991).
Constructionist scholars such as Seymour Papert (1976) have long
advocated for learning environments that incorporate education into a
``larger, richer, cultural-social experience'' (p.~7). Computers are
ideal tools for constructionist learning, providing a collective space
for self-directed exploration and skill development (Resnick, Bruckman,
and Martin, 1996). This is especially true of participatory media.
Participatory media environments are characterized by low barriers for
engagement, an emphasis on sharing, informal mentorship between
experienced members and novices, and feelings of connection among
members of the group (Jenkins, 2009: 6). Predominantly social in nature,
these environments are a space for active learning where kids develop
new skills as they create media with and for each other. Research
suggests that participating in these social communities contributes as
much to learning as technical tools like software (Kafai and Burke,
2014).

In participatory media environments, social interactions often drive the
creative and educational activities of young people. As Mimi Ito and her
co-authors (2009b) outline in \emph{Hanging Out, Messing Around, Geeking
Out,} young people socialize online in varying ways. They ``hang out''
and engage in friendship-driven activities, chatting with offline
friends and peers. They also ``geek out,'' an interest-driven activity
motivated by a shared interest or hobby. When social relationships lead
to budding interests in new activities, users participate in a
productive middle ground they call ``messing around.'' These cultural
worlds have the power to transform what one considers peers, as users
hang out in interest-driven online spaces where social relationships
lead to budding new interests. Yet, reflecting on their previous
research, Ito et al. (2018) observe that although they saw ample
evidence of social hanging out and interest-driven geeking out in their
varied field sites ``few {[}young people{]} were taking advantage of the
learning potential of digital networks'' (p.~2).

The subsequent ``connected learning'' model proposed by a team of
researchers focused on digital youth unites these activities and
identifies three crucial factors for new media and education: (1)
support of peers, (2) engagement of personal interests, and (3) ties to
academic opportunity (Ito et al., 2013). At the confluence of these
factors, young people engage with peers who share their interests to
expand and deepen their formal knowledge. Ito et al.~write that
``effective learning involves individual interest as well as social
support to overcome adversity'' (p.~4). In our work, we examine this
process to conceptualize how hanging out becomes ``figuring it
out''---as young programmers face creative problems and then use the
social features of an interest-driven sub-community to overcome them.

Education-oriented organizations, such as the Partnership for 21st
Century Skills (2019), have identified ``problem solving'' as one of the
most important and applicable skills for young people today. Students
need to develop strategies for utilizing their existing knowledge to
creatively seek solutions. In an increasingly digital world,
problem-solving skills are often acquired in online environments where
young people learn (and learn from) computer-based tools. Through these
activities they develop skills that Janet Wing (2006) has called
``computational thinking.'' Computational thinking is a way of
approaching problems that uses a broad range of concepts central to
computer science. Rather than merely the ability to use a computer,
computational thinking is based in practices and perspectives that can
be utilized in the material or digital world. Practices like dividing
things into smaller parts (modularizing), creating in test cycles
(iterating), or building on the work of others (remixing) are all
manifestations of computational thinking (Brennan and Resnick, 2012).
Our study is particularly focused on the practice of ``debugging'':
strategies employed by users to ``deal with and anticipate problems''
(Brennan and Resnick, 2012: 7).

Debugging is a key aspect of computational thinking and, by its very
nature, closely related to the idea of problem solving. In traditional
educational environments like school, students are frequently placed in
situations where they either ``get something'' or get it wrong (Papert,
1980: 21). As a computer programmer, you rarely get it right the first
time. It is through failure, and successful debugging, that programmers
learn how to problem solve and to develop strategies they can apply to
both online and offline situations. Debugging can take the shape of
experimenting with code scripts, rewriting scripts, finding examples of
scripts that work---and even more familiar debugging behaviors like
taking a break or asking someone for help (Brennan and Resnick, 2012).

Constructionist learning scholars have consistently identified
socializing as a necessary aspect of learning online. Peer communities
both motivate and support learning by providing role models, project
examples, and an audience with which to share creations. The ability to
help one another is a catalyst for social interaction and enjoyment on
game-making platforms (Bruckman, 1998). Rather than viewing programming
as a solitary activity, they argue that educational advocates would be
well served to view it as a ``shared social practice''---requiring not
just the skills to \emph{think} computationally but also to
\emph{participate} computationally (Kafai, 2016; Kafai and Burke, 2014).
Although acknowledged as entwined, becoming a better programmer and
becoming a better collaborator are predominantly conceptualized as
distinct outcomes in the constructionist literature. Learning to code is
weighed as an individual achievement, a personal outcome of acquired
skill. The social dimension is largely focused on the building of soft
skills, such as working well in teams (Kafai and Burke, 2015).

Problem solving in online spaces is often a collective endeavor. For
example, researchers studying the multi-player online game \emph{World
of Warcraft} observe that the game's capacity for social interaction
helps players to both seek and evaluate information as they jointly
develop shared understanding (Martin and Steinkuehler, 2010;
Steinkuehler and Duncan, 2008). In online software development
communities like GitHub, social features present programmers with unique
opportunities for collaborative team building and new methods of
building software (Begel, DeLine, and Zimmerman, 2010). Dabbish, Stuart,
Tsay, and Herbsleb (2012) argue that the transparent sharing of
information on GitHub allows programmers to more effectively coordinate
on projects \emph{and} better their technical skills. The social
components of these online programming communities allow groups to
advance software development through teamwork and collective learning.

Our work furthers research on collective problem solving within such
communities to consider the practices of young people and budding
programmers. Specifically, we investigate the computational practice of
debugging within the Scratch online community. In doing so, we offer a
framework for identifying and nurturing productive social interaction as
a pathway for acquiring analytic, higher-order knowledge in
youth-oriented participatory media environments.

\subsection{Empirical Setting: Scratch Online
Community}\label{empirical-setting-scratch-online-community}

Developed with young people in mind, Scratch is a visual programming
language that allows users to build interactive media by snapping
together programming blocks. Scratch's developers at the Lifelong
Kindergarten group at the MIT Media Lab situated the language within an
online \emph{community} in which young people could engage in
``construction-oriented, personally meaningful acts of creative
expression'' (Brennan et al., 2010). The social nature of the Scratch
online community is intentional: users can interact with each other
while they learn basic programming by creating online ``projects'' that
often take the form of animations and games.

Andres Monroy-Hernández (2012), creator of the Scratch online community,
explained that he sought to center socializing by building ``a space
where peers create, share, remix and even just `hang out'\,'' (p.~38).
To support this, Scratch has social features such as a ``love it''
button and enables users to remix projects, curate projects into
galleries, post on forums, and contribute comments. When users encounter
problems in constructing their projects, they often turn to comment
threads and forums to find helpful information from other users.\\
Within the large body of empirical research on Scratch, there is debate
about the relationship between socializing and learning. Qualitative
analysis and case studies by Brennan et al.~(2010), Brennan et al.
(2011), Fields et al.~(2015), Kafai and Burke (2014), and Aragon et al.
(2009) all support the argument that socializing can play an important
role in learning to program on Scratch. In a quantitative study by
Dasgupta et al.~(2016), increased commenting (used as a control
variable) was shown to have a marginal but positive effect on the number
of blocks Scratchers had in their coding vocabulary.

Other empirical work has painted a less optimistic picture about the
relationship between socializing and learning about computing in the
Scratch online community. The study from Dasgupta et al.~(2016) also
showed that increased commenting was associated with a lower likelihood
of using six complex Scratch blocks associated with computational
thinking for the first time. Gan, Hill and Dasgupta (2018) suggest that
positive feedback in the form of ``love its'' is negatively associated
with users' decisions to share subsequent Scratch projects---especially
among boys. This research reveals a tension between the community's
capacity for social activity and its capacity to promote learning
through project creation.

Brennan et al.~(2010) reflect in the concluding paragraphs of their
article \emph{Making Projects, Making Friends} that there are two
archetypes of Scratch users: ``socializers'' and more computationally
focused ``creators.'' They suggest that both of these extremes present
challenges in that ``socializers are deprived of opportunities to focus
on interactive media creation and express themselves with new forms''
while ``creators lack the benefit of learning with and from others,
developing the community's and their own capacities for creation''
(p.~78). Taken together, empirical research suggests that although
socializing on Scratch can be a pathway to computational thinking, this
potential is frequently unrealized.

\section{Data}\label{data}

The data for our study drew from the over 46 million projects and 221
million comments made by users on Scratch. Our data were collected as
part of a large research project on informal learning environments
overseen by the Committee on the Use of Humans as Experimental Subjects
at the Massachusetts Institute of Technology. Creators posted all the
projects and comments analyzed here with the knowledge that they could
be seen publicly online. Projects that have since been deleted or made
private by users have been eliminated from the dataset, respecting the
autonomy of Scratch users. Furthermore, Scratch doesn't allow real
names, requiring that users' identities remain anonymous both on Scratch
and in our data. This study only utilizes information from public data
and involved no interaction or intervention by the researchers.

\subsection{Project-Level Comment
Data}\label{project-level-comment-data}

In order to identify meaningful instances of peer feedback and begin to
understand the conditions of their occurrence on Scratch, we built a
dataset that included a random subset of 1,240 projects shared in the
community. The ``sampling unit'' was a single Scratch \emph{project}.
Our sample of projects was drawn from the population of Scratch projects
that received three or more comments---a restriction justified because
our investigation was focused on understanding collaborative problem
solving that occurs through social activity online. We felt that three
comments represented reasonable minimal activity for discussion and
socialization. The dataset included metadata about each project,
including a hyperlink to the ``live'' project on Scratch and a full
textual representation of all the comments displayed in chronological
order but threaded in a way that is identical to how it would have been
displayed on Scratch.

\subsection{User Project History Comment
Data}\label{user-project-history-comment-data}

Because one of our goals was to understand the development of
collaborative problem-solving skills, we constructed a second dataset of
users' experiences over time. We hoped that these data would reveal the
processes through which users do or do not develop computational
thinking skills. To build an appropriate dataset, we wrote a Python
script to collect \emph{all} the commenting activity on projects that
were created by 53 users who we selected using the methodology described
in the next section. We created a separate file for each of the users
and included the comments made on every published project created by
each of them during their time on Scratch. Although 53 is not an
enormous number of users, it reflects a large dataset because each user
can share many projects and each project can receive many comments. The
number of projects shared by these users ranged from one to more than
1,500 with a median of 29 projects. Like activity in Scratch in general,
the distribution of activity across users is highly skewed. In total,
the user trajectories comprised a dataset of 3,779 comments left on
5,213 projects.

\section{Methodology}\label{methodology}

Our analysis was conducted in three stages. Stage I consisted of an
inductive analysis of project-level comment data that led to the
identification of a construct that we call \emph{participatory
debugging}. Stage II was a simple content analysis conducted on a random
sample of Scratch projects that established how common participatory
debugging is on Scratch. Stage III was a second inductive analysis that
considered Scratch users' project histories of shared projects to
identify a set of potential factors for the emergence of participatory
debugging.

Because the goals of Stages I and III involve identifying recurrent
characteristics of social interactions, they are well suited to a
grounded theory (Glaser and Strauss, 1967). Grounded theory is an
iterative procedure, which utilizes the ``constant comparative method''
(Glaser and Strauss, 1967: 101). Constant comparison begins with coding
data into as many categories of analysis as possible, during a period of
``open coding'' (Lindlof and Taylor, 2011: 246). Memos are then used to
create a codebook, which contains definitions of the codes, their
relation to one another in thematic categories, and examples from the
text (Lindlof and Taylor, 2011: 251). As new categories and codes are
revealed, they shape and are applied to subsequent and previous data
(Glaser and Strauss, 1967: 109).

Stage I consisted of a grounded theory analysis of 640 projects and
their full comment threads conducted in parallel by two authors. The
result of Stage I was a memo written by the first author that described
the construct of \emph{participatory debugging}. The content of this
memo is described in the first portion of the Findings section and was
used to guide subsequent stages of our analysis. Stage II included a
content analysis of 600 randomly selected Scratch project comment
threads in which we sought to identify the presence or absence of
participatory debugging activity as described in the Stage I memo.
Following Neuendorf (2017), our content analysis involved a ``norming''
step in which 100 projects were coded for the presence/absence of
participatory debugging by both authors and any disagreements were
discussed (stage II-A). Following this step, 300 projects were coded by
two coders and intercoder reliability was assessed (II-B). After
establishing reliability, an additional 200 projects were coded by one
of the authors (II-C).

Although the random samples of projects used in Stage I and II provided
examples of participatory debugging and established how widespread it
is, we found it difficult to understand the social contexts that
supported the emergence of the practice. Stage III sought to address
this by conducting a second grounded theory analysis of all projects
shared by particular users over time and the ``cultures'' that emerged
in their comment threads. The user-level dataset used in Stage III
included all projects (and the projects' associated comments) shared by
the 53 users who we identified as showing evidence of participatory
debugging in Stage II. Importantly, this meant that each user project
history had at least one instance of socially situated problem solving.
The dataset used in Stage III included 5,213 projects and 3,779
comments. The user histories were randomly split into two sets, each
coded by one of the authors. Exemplary or difficult cases were
highlighted, shared, and discussed. In all cases, examples of comments
and drafts of memos were read and discussed by the full research team
throughout the iterative process of theory generation. The datasets were
analyzed with a focus on ongoing interactions that represented and
shaped participatory debugging. We sought to understand the processes
between people, situations, and events and how they influence each
other---an approach known as process theory. Process theory deemphasizes
individual participants in favor of understanding ``the particular
context within which participants act and the influence that this
context has on their actions'' (Maxwell, 2013: 30).

Table 1 lays out the stages and steps of our methodology within stages
in terms of the number of projects, number of comments, and the number
of coders involved. In total, this work is based on analysis of over
14,733 comments left on 6,453 projects.

\begin{longtable}[]{@{}lllll@{}}
\toprule\noalign{}
\endhead
\bottomrule\noalign{}
\endlastfoot
Table 1 & & & & \\
Qualitative data analysis steps & & & & \\
\ul{Stage} & \ul{Description} & \ul{Projects} & \ul{Comments} &
\ul{Coders} \\
I & Construct Definition & 640 & 2,846 & 2 \\
II-A & Norming & 100 & 1,168 & 2 \\
II-B & Reliability Test & 300 & 3,989 & 2 \\
II-C & Additional Projects & 200 & 2,951 & 1 \\
III & User Project Histories (53 users) & 5,213 & 3,779 & 1 \\
TOTAL & & 6,543 & 14,733 & \\
\end{longtable}

\section{Findings}\label{findings}

Our findings from Stage I are the construct of \emph{participatory
debugging}. Our findings from Stage II are the results of a content
analysis that describe how common participatory debugging is on Scratch.
Our findings from Stage III include three conditions in which
participatory debugging is performed most effectively. We present each
in turn.

\subsection{Stage I: Participatory Debugging
Practices}\label{stage-i-participatory-debugging-practices}

The most salient and important theme to emerge from the grounded theory
analysis of 640 projects in Stage I was a theme we ultimately labeled
\emph{participatory debugging.} Participatory debugging describes a
computational practice in which users leverage social interaction in
order to overcome challenges and achieve their goals. In the most basic
sense, participatory debugging occurs when the creator of a project
identifies a problem with their own project and communicates this fact
to others explicitly. This often happened in the project description
when a project creator would acknowledge shortcomings (``I know it has
some errors'') or give specific warnings (``the button doesn't work'').
In these situations, project creators are demonstrating their awareness
of how engaging with a larger online community is an important first
step toward receiving help---and that other Scratchers who view the
project (thus experiencing the error) might also possess the knowledge
to fix it. When a creator identifies a glitch in their own project, they
open space for conversations and comments that lead to the other
problem-solving behaviors that produce a solution.

Project creators may also ask for help or assistance explicitly. Rather
than requesting help from a single user, creators most often addressed
the larger community---issuing open-ended questions or declarations of
distress. Requests can range from being general (``I need help!'') to
more specific (``how do you make a cloud list?'') In response, other
Scratchers may comment with explanatory solutions to the problem: ``to
improve the jumping use this script\ldots{}'' or ``if you want to reduce
the forever loop, you can\ldots{}''

In the most basic manifestation of participatory debugging, a Scratcher
may simply state that a glitch is present in the project they're viewing
with comments like ``it glitches'' or ``it's broken.'' Scratchers may
also identify \emph{how} the project is glitching (``the commands don't
do anything'') or, more helpfully, what is causing the glitch and how to
fix it: ``great! instead of using {[}when (key) key pressed{]} you might
want to use {[}forever {[}if {[}key (key) pressed?{]}{]}{]}... it will
make the movement smoother!'' Even when creators don't identify problems
in their own projects, other Scratchers identify problems and provide
solutions without being solicited. The appearance of these conversations
demonstrates a norm of participation in which Scratchers see others and
themselves as a resource for improvement.

Importantly, participatory debugging is not merely an activity that
occurs between a project creator and a single expert user. Instead, it
mobilizes the resources of multiple users to fix multiple problems on
multiple projects, within the Comments space of a single project. In
what follows, we see a conversation between four Scratchers, none of
whom are the project creator. The first two comments both identify a
glitch (the speed of the text), with the first Scratcher providing a
solution. The third Scratcher then asks a seemingly unrelated
programming question, receiving a detailed response from a fourth
telling him what to do---add an additional programming block---and how
that block works in relation to other blocks:

\begin{quote}
P1: you should make it so that once somebody is done reading the text,
they can press space to hear the next thing instead of the Scratcher
having to rush to read it.

P2: Text is too fast... And your spelling is not that good... Sorry.

P3: HeLp! There is this button in Control which says "When I receive
\_\_\_\_\_\_" I typed in something...but I still do not understand what
the message means. HELP! (PS. How do I work the message?)

P4 {[}in response to P3{]}: You have to use the broadcast block. Think
of it this way, the broadcast block makes the sprite say something all
the other sprites can hear. The when I receive block really means when I
hear. If that doesn\textquotesingle t help comment on one of my projects
and I can do more explaining.
\end{quote}

When Scratchers are faced with a problem, they seek out the successful
projects of others, leaving comments for more experienced Scratchers.
They may comment to explain their situation and to elicit an answer, or
they may connect their comments to the project they are viewing---asking
``how did you do that?''

This manifests in unique ways, because the Scratch online community
allows users to ``see inside'' an animation to view how it is coded. The
ability to view the code of a project allows the project to serve as a
demonstration of successful coding concepts. Even though the coding
blocks are always visible to other Scratchers, social engagement through
the Comments space still serves as a valuable explanatory tool. For
example, in an outer space themed project a Scratcher asks ``Can you
please explain how you got the position of the planet'' and the creator
responds ``Ok, so first you make the sun. Where ever he is, make a
variable for his x and y position \ldots.'' and then continues to
explain, step by step, how to execute the project.

With its focus on computational practices, participatory debugging
describes problem-solving behaviors concerned with the functionality of
a project, such as identifying glitches or problems with code, rather
than those that were more aesthetic suggestions related to the way a
game looks or how it is played. Although distinct, we found that
examples of these behaviors have a high co-occurrence with participatory
debugging: projects in which users discuss how a game \emph{looks} are
also likely to discuss how a game \emph{works}. In both cases, project
creators and Scratchers engage with each other as a social resource.

\subsection{Stage II: How Common Is Participatory
Debugging?}\label{stage-ii-how-common-is-participatory-debugging}

In Stage II, we conducted a content analysis of Scratch projects in
order to determine how common participatory debugging is on Scratch. Our
content analysis followed the methodology laid out by Neuendorf (2017)
and involved human coders viewing projects and comment threads and
judging whether the threads included examples of participatory debugging
as we defined it in the memo that resulted from Stage I (summarized in
depth in the previous section). As is typical in content analysis, our
approach relied on the judgment of our human coders and did not require
the presence of particular keywords. To assess intercoder reliability of
the construct, we computed Krippendorff's alpha on the sample of 300
independently coded projects conducted in Stage II-B and found that we
were able to reliably code projects for the presence or absence of
participatory debugging (\(\alpha\)=0.79).

Within the sample of 600 projects coded in Stage II, we identified 53
examples of participatory debugging or 8.8\% of the projects in our
sample. Using a binomial ``exact'' calculation, our 95\% confidence
interval for the true value in the population from which our sample was
drawn---i.e., projects with at least 3 comments---is estimated as
between 6.7\% and 11.5\%. Although both our team and the Scratch design
team were surprised that the number was this high, it is evidence that
participatory debugging practices are rare on Scratch.

\subsection{Stage III: Social Situations and Commenting
Culture}\label{stage-iii-social-situations-and-commenting-culture}

In the final step of our analysis, we sought to understand how and when
participatory debugging was fostered in group-oriented cultures of
problem solving. To do so, we conducted an additional round of grounded
theory analysis on all the projects shared by the 53 users who had
showed evidence of participatory debugging in Stage II. We found that
online commenting spaces allow for participation in the form of deep
social engagement involving shared interests or personal feelings as
well as discussions related to programming. Although the technical
qualities of these social environments---the ability to comment, the
ability to remix---were consistent between users, how they were used was
not. It is not the mere existence of commenting spaces that leads to
participatory debugging, but rather specific cultures of commenting. We
identified three factors that facilitated or hindered participatory
debugging---from the most basic to the more complex: (1) a sustained
community, (2) identified problems, and (3) ``porous'' conversations
that aren't intensely focused on a singular topic.

\textbf{Sustained community.} A sustained community is defined by two
characteristics: the presence of others and consistent engagement. While
the presence of others may seem like an obvious aspect of participatory
debugging, it is not a given on Scratch that other users will comment on
a creator's project. For example, in one user's history we see a series
of eight projects (all revised versions of the same project) in which he
asks each time: ``Please post suggestions for full game,'' but he gets
no response. Overall, this user has a very low occurrence of comments
throughout his projects. Without the engagement of others, his calls for
help do not result in any real dialogue or opportunity for learning. A
very similar call for feedback is met with a response and a solution
when it came from another user who regularly receives comments from
about a dozen Scratchers. When this second user asks ``is anyone else
having slider difficulties,'' another Scratcher responds with the exact
spot in the code that is broken. The creator replies with a detailed
explanation of the code: ``Yeah, I can\textquotesingle t really fix that
because here\textquotesingle s why: repeat until \textless(answer)
\textgreater{} {[}0{]}\textgreater{} ask {[}Hour{]} \ldots. It will skip
past the minute part because it already satisfies the sensor blocks''
and the other Scratcher replies again, helping to rewrite it.

Some groups of users interact repeatedly in ways that create a sense of
social momentum. These users tend to interact across a given user's
projects in both an active and a sustained way. This ongoing interaction
is particularly important in leading to engagement and, eventually, to
fixes to code-based issues in projects. These users have lively comment
streams, characterized by reciprocity and lending support. The energy of
the conversations seems to attract even more users. As these groups
grow, participatory debugging practices flourish.

Although participatory debugging is not possible in the absence of other
users, previous work suggests that intense sociality and popularity can
often be at odds with subsequent engagement (Gan et al., 2018) and
learning (Brennan et al., 2010). Sustained communities of users often
amplify the visibility of members' work by adding projects to
user-curated galleries, leading to an increase in project comments and
the solidification of a sustained community.

Having a small group of users that cares about one another's
projects---the kind of users who would undertake activities like placing
each other's projects in galleries---creates the potential for
commenting aimed at code-oriented problem solving. It is the presence of
this shared community, and not necessarily the high number of comments,
that most strongly contributes to ``figuring it out.'' In fact, some of
the most highly commented upon projects had neither evidence of problem
solving nor evidence of sustained group interaction. For example,
projects that were featured on Scratch's homepage received many positive
comments from many users, and projects that were part of more topically
focused fan groups had extensive commenting histories---but these
comment volumes didn't support the emergence of participatory debugging.

\textbf{Identified Problems.} Projects that have identified problems
encourage substantive feedback from other users. Substantive feedback
contains information and insight, in contrast to the more generic forms
of praise (``cool project!'' or ``Nice work'') that fill the Comments
space on Scratch. Substantive feedback occurs when users comment
directly about particular aspects of a project or technique, involving
phrases like ``it would be better if\ldots{}'' or ``I think you
should\ldots{}'' or ``you need to\ldots{}'' This feedback can either be
unsolicited (i.e., a suggestion offered unprompted) or in response to a
question asking how a particular coding technique might be achieved.
When substantive feedback occurs, it tends to be more likely that a
group of users will engage in participatory debugging.

Substantive feedback is facilitated through a Scratcher's ability to
recognize problems within projects. This is true of project creators who
identify problems in their own projects and of visitors who identify
problems through their interactions with others' projects. This can
occur when a creator indicates that something in their project isn't
working or a visitor states that there's a bug. Even when Scratchers do
not have the expert language to specify what the problem is or how to
solve it, the interactive nature of projects on Scratch allows other
visitors to experience the glitch for themselves and then view inside
the project to investigate.

A key feature of Scratch is the ability to ``remix'' a project. Users
turned to this feature to go beyond textual explanations and to
demonstrate how a project may be fixed. In the example below, a project
creator is attempting to build a script that would add viewers of the
project to a continually updated list. However, the project needs a
block that draws on Scratch data stored in the cloud---a feature that is
only available to more experienced Scratchers. The visitor uses the
remix function to demonstrate a workaround, ``decoding'' the functioning
of the list block:

\begin{quote}
P1: Hmmm...unfortunately, this won\textquotesingle t work because it
isn\textquotesingle t a cloud list...I\textquotesingle ll post an
example on how you WOULD make it work...:D

CREATOR: oh. how do you make a cloud list? there isn\textquotesingle t
any option to

P1: You can\textquotesingle t...yet. :P You have to use a system of
decoding cloud variables to do it...hold on...I\textquotesingle m almost
done...:P

P1: Alright here it is
\end{quote}

Scratch exemplifies a learning community that doesn't require expertise
to get expert help.

Rather, Scratchers can identify that \emph{something} is wrong, and
other Scratchers can use the social features built into Scratch---such
as viewing the inside of a project or remixing---to provide substantive
feedback.

\textbf{Low Topic Density (``Topic Porousness'').} Finally, it was clear
from our user project history data that neither the volume of users nor
the presence of identified problems was a sure sign of participatory
debugging. Groups whose interactions are heavily focused on a particular
topic (such as role-playing games, Sailor Moon, or other fandoms) often
have active, extensive comment streams involving multiple users, but
they don't often exhibit computational thinking practices. All or most
interactions in these groups are focused on the topic of interest rather
than on improving the project. Furthermore, even when problem solving
does occur (a user asking for help, for example) it is drowned out by
the presence of fan-related conversations. The context and content of
discussions matters in the quest for participatory debugging. For
example:

\begin{quote}
P5: the pictures dont go smooth enough, not enough images, it skips
right from the start to part way through.

P6: i love the melancholy of haruhi suzumiya

P7: I LUV THE MELANCHOLY OF HARUHI SUZUMIYA!!!
\end{quote}

This discussion can be compared to more porous conversations, which are
less focused on the abstract theme of the game and have space for
discussions about the game itself: how it works, how it is played, and
how it looks.

However, there is an implicit tension here. Involvement in fan
sub-communities is a great way to connect to other users, especially
those who are highly active and have wide reaching networks that can be
a resource for achieving the presence and sustained engagement of
others. However, the Scratchers who participate in these sub-communities
utilize the commenting space to ``geek out.'' When identifiable problems
are introduced into these topic-dense social spaces, they easily become
lost in crowded conversations.

For this reason, we identify openness to broad conversation---what we
refer to as ``topic porousness''---as being an essential aspect of
participatory debugging. We use the term \emph{porousness}
metaphorically: in the same way porous physical materials allow
substances to easily pass through them, porous topics are those that can
easily become permeated with or joined by other types of conversations.
Rather than a measure of the number of topics a group covers (an idea
like breadth), porousness describes a kind of openness and
responsiveness of commenting cultures. Porous commenting cultures are
not so dense that new topics remain barricaded from the flow of
conversation. Porous conversations allow newly presented ideas to
transform the conversation at hand and become salient to other users.
These commenting cultures can contain not only discussions of fandom but
also conversations about programming and even, like many of the Scratch
projects we read, heartfelt sharing of personal lives and beliefs. These
conversation spaces represent the highest hopes for constructivist
learning scholars---as meaningful worlds give rise to meaningful
learning. Our analysis reveals that balancing interest-driven engagement
with incentives for learning is a core challenge on Scratch.

\section{Discussion: Supporting Participatory
Debugging}\label{discussion-supporting-participatory-debugging}

The results in Stage II suggest that participatory debugging is rare on
Scratch. Our explanation for the infrequency of participatory debugging
is that each of the factors identified above serve as essential, but
insufficient, ingredients for the practice's emergence. The presence of
all three of the factors---a sustained community, identified problems,
and porous conversation---are necessary to maximize the possibility that
Scratch's social features will contribute to learning. For example, we
found many examples of commenting cultures with a sustained community
that never engaged in participatory debugging because they were missing
one or both of the other factors. A simple diagram is provided in Figure
1 to visualize all seven possible combinations of the three factors.
Although the richest opportunities for learning identified in our data
would be in the center of the diagram marked by the star, examples of
all seven combinations were present in our data. Commenting cultures
with two of the three features (the numbered areas in the figure)
reflect the biggest opportunities for supporting participatory
debugging, and thus are the most productive sites for design
interventions. Yet each combination presents unique challenges. We
discuss the three combinations in turn.

\begin{figure}
\centering
\includegraphics{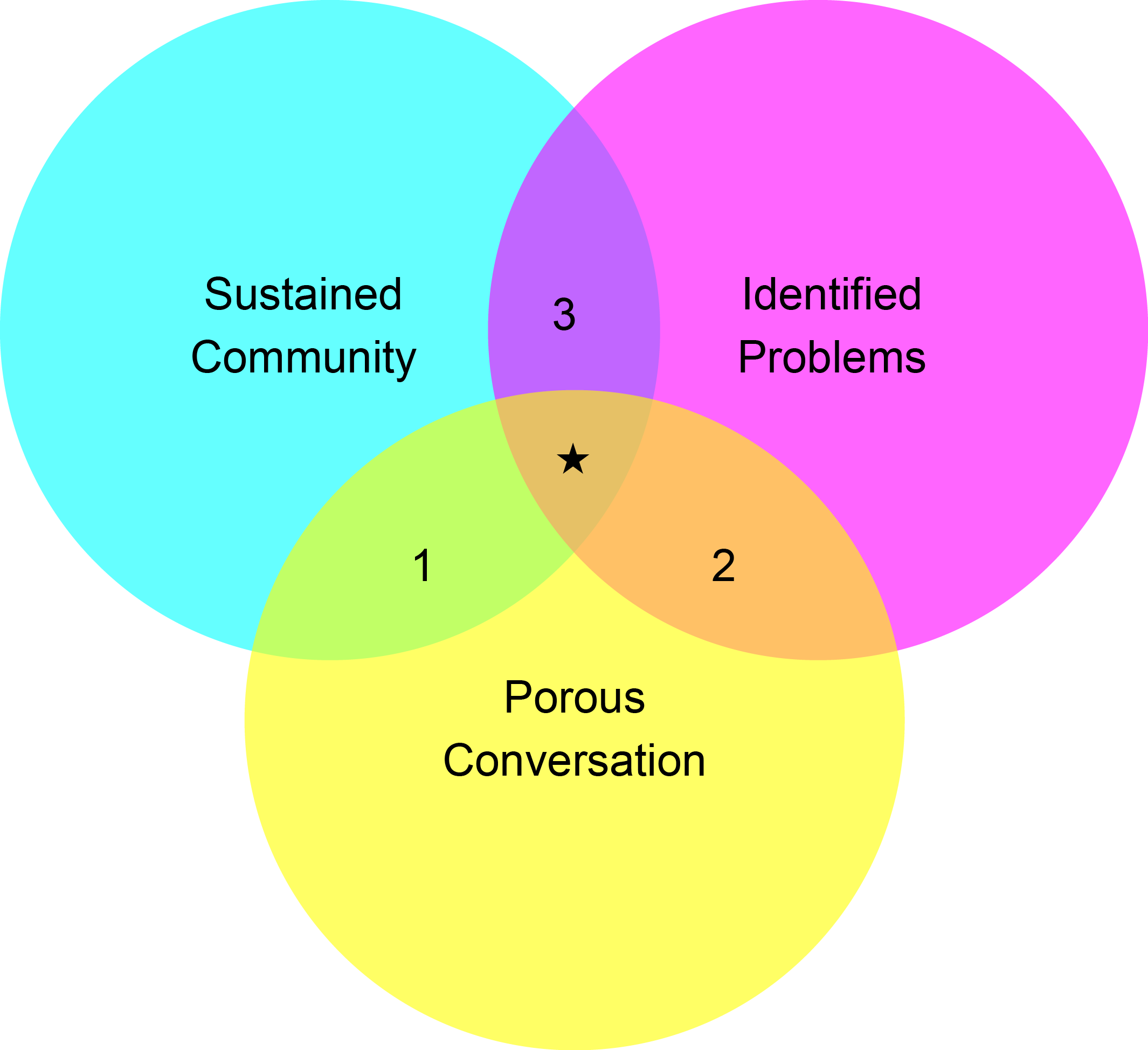}
\caption{Venn diagram visualizing the relationship between the three
social factors that foster participatory debugging.}
\end{figure}

In Area 1, we found examples of commenting cultures that are
characterized by a sustained community and porous conversations. These
are cultures that are highly social and porous but where users never
identified problems or asked for help. Despite listening ears and
wide-ranging interactions, they don't collectively share knowledge to
overcome problems. Groups of users in this category have the greatest
potential for participatory debugging in that their openness and
engagement mean that an attempt at problem solving will find fertile
ground. Designers of online communities like Scratch could promote
learning by prompting creators in these environments to reflect on their
challenges publicly in order to initiate problem solving in ongoing
conversations.

In Area 2, we identified commenting cultures that have well-defined
problems and porous conversations but lack a sustained community. Here,
project creators are aware of problems and open to fixing them, but
there simply aren't other Scratchers present to help them. Environments
like Scratch can encourage learning by promoting connections among users
in these situations. To engage in participatory debugging, users need to
be able to find and attract users who are willing and able to jointly
seek a solution. A ``needs help'' tool that makes struggling projects
more widely visible---for example, collecting them into a ``needs help''
gallery---could help build a sustained community for projects in these
settings.

In Area 3, there are commenting cultures that have both sustained
communities and identified problems but had high topic density. When the
topic is programming, these cultures contain some of the most inspiring
instances of participatory debugging. Here, communities of technically
knowledgeable users come together to push one another to achieve and
learn complex coding routines. That said, most of the groups in our
dataset with high topic density ran the risk of drowning out attempts at
problem solving in favor of more thematic conversations based in fandom
or role playing. Although fandom sub-communities are their own kind of
geeking out, they aren't generally geeking out about coding or
computers. This does not mean that fandom-focused Scratch
sub-communities are not learning computational thinking concepts in
order to create and build fandom-oriented projects. In our data,
however, they typically are not learning computational practices as part
of their social interaction. Even when problem solving is introduced as
a topic, it is too easy to overlook because users focus on a more
thematic conversation. Although supporting participatory debugging in
these contexts is challenging, a ``needs technical help'' flag attached
to comments focused on solving a problem could aid in bringing attention
to the identified problem and in breaking open dense conversations.

The tension between personal interests and more focused productive
activity represents a widely observed but previously under-theorized
trade-off in participatory learning environments. Essential to
constructionist modes of learning is a dedication to acquiring knowledge
for a ``recognizable personal purpose'' (Papert, 1980: 21). On Scratch,
this means building projects rather than rote memorization of
programming concepts. Learning extends beyond the impersonal and the
abstract, becoming expressive and personally meaningful (Resnick,
Bruckman and Martin, 1996). Thematic interests in popular culture often
motivate project design and connect users in social communities. For
example, in her micro-analysis of MOOSE crossing, Bruckman identifies
``significant shared interests'' (in this case, \emph{Star Trek}) as
building bonds between and strongly contributing to their integration
into the community (p.~49). The themes, stories, and content of projects
are also especially important on Scratch (Kafai and Burke, 2014: 114).
In our analysis, projects that lack a sustained commenting culture may
be evidence of project creators who are not tapping into the power of
interest-driven communities. However, our analysis also suggests that
the energy present in communities that assemble around shared thematic
interests needs to be balanced by tools that can aid in
programming-focused interactions.

While our analysis discusses the three factors as static ingredients of
commenting cultures conducive to participatory debugging, we found that
they often exist as part of a complex network of feedback loops. For
example, identified problems could inspire sustained conversations as
groups of users repeatedly interact to test, iterate, and determine
solutions to problems. Alternatively, porous conversations that contain
personal, interest-driven, and computationally oriented topics could
also promote a sustained community as users develop meaningful
connections to one another. As a result, we found that the rare
confluence of the three factors would often support participatory
debugging not just once but repeatedly over time within a given user's
projects. Although the presence of participatory debugging was
relatively rare, it can become a repeated feature of a user's activity
on Scratch when commenting cultures exhibit these three features.

\section{Conclusion}\label{conclusion}

Glitches, bugs, and errors are natural parts of programming, but for
individuals with entrenched negative beliefs about their abilities,
these initial failures become proof that they can't do it (Papert,
1980). The impact of initial failure may be especially strong for groups
of people, such as girls, who are widely stereotyped as being inherently
bad at math and activities, like computing, associated with mathematical
skills (Spencer, Steele, and Quinn, 1999). In \emph{Mindstorms}, Papert
(1980) observes that only rare, exceptional events allow people to
overcome their deeply held notions about their own capabilities.
Creating projects on Scratch is an opportunity for these events. By
studying the social process through which users problem solve, we hope
to better understand the spaces through which people learn to overcome
failure in the pursuit of learning.

Our project indicates that young people do indeed turn to their social
resources in order to overcome challenges and solve problems. That said,
among the creators of these projects, we find very different
trajectories and many examples of users who do not develop and hone
higher-order computational thinking practices. A sustained community,
identified problems, and porous conversation are all important factors
of participatory debugging. Supporting learning online requires that
creators of such environments balance the social resources that exist in
interest-driven sub-communities with tools that can promote the learning
of concepts and analytic thinking skills. The connected learning agenda
seeks to link ``deep vertical expertise'' (like that which exists in
interest-driven communities) with practices that are recognized as a
source of professional opportunity (like programming) (Ito et al, 2013:
56). This link possesses what Ito and her research partners (2018)
identify as the as of yet unrealized potential of connected learning (p.
170).

We hope that further research will show that the dynamics we have
described extend far beyond programming communities like Scratch. We
believe that the participatory debugging practices we identified are
just one example of important forms of computer-mediated social
problem-solving practices enabled by new media. Although further work is
necessary to establish the generalizability of our findings, we believe
that the social of participatory debugging are features of healthy
online problem-solving cultures more generally. We believe that our
findings and the framework we have identified will have parallels in a
broad range of contexts engaged in supporting connected learning. Our
work shows that the literature on participatory media is ready to move
beyond discussions of whether youth socializing can or cannot lead to
the development of higher-order computational thinking skills. We
present our work as a productive first step in asking what is required
to effectively support and encourage both learning and social
interaction.

\section{Acknowledgments}\label{acknowledgments}

We would like to acknowledge and thank members of the Scratch online
community for inspiring this work through their creative problem
solving. We are deeply grateful to Sayamindu who worked with us to help
construct the dataset used in this paper, offered pointers to relevant
work, and who provided feedback and support throughout the project. We
would also like to thank Mitchel Resnick, Natalie Rusk, and the editor
and anonymous reviewers at New Media and Society for their thoughtful
feedback and support. We presented early versions of this research to
the Community Data Science Collective and the International
Communication Association's section on Children, Adolescents and the
Media and are are grateful to members of both groups for their valuable
feedback and suggestions. Financial support for this work came from the
National Science Foundation (grants DRL-1417663 and DRL-1417952).

\section{References}\label{references}

Aragon CR, Poon SS, Monroy-Hernández A, and Aragon D (2009) A tale of
two online communities: Fostering collaboration and creativity in
scientists and children. \emph{Proceedings of the Seventh ACM Conference
on Creativity and Cognition}: 9--18.
https://doi.org/10.1145/1640233.1640239

Begel A, DeLine R, and Zimmermann T (2010) Social media for software
engineering. \emph{Proceedings of the FSE/SDP Workshop on Future of
Software Engineering Research}: 33--38.
https://doi.org/10.1145/1882362.1882370

Brennan K, Monroy-Hernández A and Resnick, M (2010). Making projects,
making friends: Online community as catalyst for interactive media
creation. \emph{New Directions for Youth Development}, \emph{2010}(128):
75--83. https://doi.org/10.1002/yd.377

Brennan K and Resnick M (2012). \emph{New frameworks for studying and
assessing the development of computational thinking}. In: Educational
Research Association, Vancouver, Canada.

Brennan K, Valverde A, Prempeh J, Roque R and Chung M (2011). \emph{More
than code: The significance of social interactions in young people's
development as interactive media creators}. In: ED-MEDIA, Lisbon,
Portugal.

Bruckman A (1998) Community support for constructionist learning.
\emph{Proceedings of the ACM Conference on Computer Supported
Cooperative Work,} pp.~47--86. https://doi.org/10.1023/A:1008684120893

Dabbish, L, Stuart C, Tsay J, and Herbsleb J (2012) Social coding in
GitHub: Transparency and collaboration in an open software repository.
\emph{Proceedings of the ACM 2012 Conference on Computer Supported
Cooperative Work}, pp.~1277--1286.
https://doi.org/10.1145/2145204.2145396

Dasgupta S, Hale W, Monroy-Hernández A, and Hill BM (2016) Remixing as a
pathway to computational thinking. \emph{Proceedings of the 2016 ACM
Conference on Computer-Supported Cooperative Work \& Social Computing},
pp.~1438--1449. https://doi.org/10.1145/2818048.2819984

Fields DA, Pantic K, Kafai YB (2015) ``I have a tutorial for this'': the
language of online peer support in the scratch programming community.
\emph{Proceedings of the 14\textsuperscript{th} International Conference
on Interaction Design and Children,} pp.~229-238.
https://doi.org/10.1145/2771839.2771863

Gan EF, Hill BM and Dasgupta S. (2018). Gender, feedback, and learners'
decisions to share their creative computing projects. \emph{Proceedings
of the ACM Conference on Computer Supported Cooperative Work,} pp.
54:1--54:23. https://doi.org/10.1145/3274323

Glaser BG and Strauss AL (1967). \emph{The discovery of grounded theory:
strategies for qualitative research}. Chicago: Aldine Publishing
Company.

Ito M, Baumer S, Bittanti M, boyd d, Cody R, Herr-Stephenson B, Horst H,
Lange P, Mahenran D, Martínez K, Pascoe C, Perkel D, Robinson L, Sims C
and Tripp L (2009b) \emph{Hanging out, messing around, and geeking out:
kids living and learning with new media}. Cambridge, MA: MIT Press.

Ito M, Gutiérrez K, Livingstone S, Penuel B, Rhodes J, Salen K.,
\ldots{} Watkins SC (2013) \emph{Connected learning}. Irvine, CA:
Digital Media and Learning Research Hub.

Ito M, Martin C, Pfister RC, Rafalow MH, Salen K and Wortman A (2018)
\emph{Affinity online: How connection and shared interest fuel
learning}. New York: NYU Press.

Jenkins H (2009) \emph{Confronting the challenges of participatory
culture: media education for the 21st century}. Cambridge, MA: The MIT
Press.

Kafai YB (2016, August) From computational thinking to computational
participation in K--12 education. \emph{Communications of the ACM},
\emph{59}(8).

Kafai YB \& Burke Q (2014) \emph{Connected code: Why children need to
learn programming.} Cambridge, MA: MIT Press.

Kafai YB and Burke Q (2015) Constructionist gaming: Understanding the
benefits of making games for learning. \emph{Educational Psychologist},
\emph{50}(4): 313--334. https://doi.org/10.1080/00461520.2015.1124022

Lave J \& Wenger E (1991) \emph{Situated learning: Legitimate peripheral
participation} (1st ed.) Cambridge, England: Cambridge University Press.

Lindlof TR and Taylor BC (2011) \emph{Qualitative communication research
methods} (3rd ed.) Thousand Oaks, CA: SAGE.

Martin C and Steinkuehler C (2010) Collective information literacy in
massively multiplayer online games. \emph{E-Learning and Digital Media},
\emph{7}(4): 355--365. https://doi.org/10.2304/elea.2010.7.4.355

Monroy-Hernández, A. (2012). \emph{Designing for remixing: Supporting an
online community of amateur creators}. Thesis, Massachusetts Institute
of Technology, Cambridge, MA. Retrieved from
http://dspace.mit.edu/handle/1721.1/78202

Maxwell JA (2013) \emph{Qualitative research design: an interactive
approach} (3rd ed.). Thousand Oaks, CA: SAGE.

Neuendorf, KA (2017) \emph{The content analysis guidebook} (2nd ed.).
Thousand Oaks, CA: SAGE.

Papert S (1976) \emph{Some poetic and social criteria for education
design}. Retrieved from http://dspace.mit.edu/handle/1721.1/6250

Papert S (1980) \emph{Mindstorms: Children, computers, and powerful
ideas}. New York, NY: Basic Books.

Partnership for 21st Century Skills (2019) \emph{Framework for 21st
century learning}. Report. Retrieved from
http://www.p21.org/about-us/p21-framework/260

Resnick M, Bruckman A, \& Martin F (1996) Pianos not stereos: creating
computational construction kits. \emph{Interactions}, \emph{3}(5):
40--50.

Steinkuehler C and Duncan S (2008) Scientific habits of mind in virtual
worlds. \emph{Journal of Science Education and Technology},
\emph{17}(6): 530--543. https://doi.org/10.1007/s10956-008-9120-8

Spencer SJ, Steele CM and Quinn DM (1999). Stereotype threat and women's
math performance. \emph{Journal of Experimental Social Psychology
Journal of Experimental Social Psychology}, \emph{35}(1): 4--28.

Wing JM (2006) Computational Thinking. \emph{Communications of the ACM},
\emph{49}(3): 33--35. https://doi.org/10.1145/1118178.1118215

\end{document}